\documentclass[a4paper,11pt]{article}
\usepackage{jcappub} 
\usepackage{lineno}

\title{\boldmath A New Way to Detect Axions from $\rm{A\bar{Q}Ns}$ Captured in the Earth}

\author[1]{Ionel Lazanu} 
\author[2]{and Konstantin Zioutas} 
\affiliation[1]{University of Bucharest, Faculty of Physics\\
POBox MG-11, Magurele, Ilfov Romania}
\affiliation[2]{University of Patras, Physics Department\\
26504 Patras-Rio, Greece}

\emailAdd{ionel.lazanu@g.unibuc.ro}

\emailAdd{ zioutas@cern.ch}

\abstract{Macroscopic dark matter with dominating strong interactions, supposed to be composites, represents an alternative to the most popular WIMP particles. Predicted in various models as strangelets, nuclearites, nuggets, having different internal structures and properties, but not yet observed experimentally, these forms of dark matter are associated with the existence of a large number of still unexplained observations. Nuggets, initially predicted by Witten, were reconsidered from the point of view of their internal structure and further theorized in 2003 by Zhitnitsky as axion quark nuggets and axion antiquark nuggets, as being made of quarks in a superconducting colour state, in the core, an electrosphere of electrons or positrons and a domain wall that maintain the stability of the macros with an incredible density, mass in the gram range and radius on the order of micrometers.
If the existence of $\rm{AQNs}$ and $\rm{A\bar{Q}Ns}$ is demonstrated, two major open problems in physics could be addressed simultaneously: they would constitute viable dark matter candidates and, at the same time, provide a natural mechanism for restoring matter–antimatter symmetry in the Universe.
The experimental evidence of the $\rm{AQNs}$ and $\rm{A\bar{Q}Ns}$ is a challenge for current and future experiments. The present study demonstrates that if these macroscopic systems exist, axions produced by $\rm{A\bar{Q}N}$s could be detected by the next generation of neutrino physics experiments using liquid noble gases, due to their huge active volumes.}

\begin{document}
\maketitle
\flushbottom

\section{Introduction}
\label{sec:intro}
Recently, two papers \cite{Cappiello:2025yfe} and  \cite{Alam:2025qbq}, proposed the idea that if dark matter has a nonzero scattering cross section with the nucleons or electrons, then it should collide with different astrophysical objects and can be captured. This process will be continued inside the star or planet. Because the planets in our solar system are approximately 4.5 billion years old, it can be supposed, in the Earth particular case, that DM gradually sinks toward the Earth's center and will finally concentrate in the core contributing to the increase of the density. More sequences of processes can be considered; for example: a) After the penetration of DM inside the celestial body the scattering processes will continue with a decrease in their kinetic energies and finally will be accumulated in the central region; b) Dark matter begins to annihilate. Some of the products can escape directly from the core of the Earth, or from decay chains in stable or to long-lived particles or mediators and can be detected. The expected signals produced from these processes can be detected as a new way to highlight the indirect contributions from the dark matter; c) In the sequence of interactions inside the Earth or as effect of annihilations in all the cases when charged particles or photons are produced, finally the energy will be converted in heat. 
The Earth's Structure \cite{BethGeiger} is composed of four distinct layers. Starting from the center, these are: the inner core, the outer core, the mantle and the crust.  The inner core is solid metal ball, extremely dense, with a radius of 1220 kilometers. It is made mostly of iron and nickel. The inner core spins a bit faster than the rest of the planet. Their temperature is about $5400^{\circ}\,\rm{C}$. The outer core is also made from iron and nickel, just in liquid form, with a dimension of 2200 km. Heated largely by the radioactive decay of the elements uranium and thorium, this liquid churns in huge, turbulent currents. That motion generates electrical currents. They, in turn, generate Earth’s magnetic field. The mantle is close to 3000 kilometers with a composition mostly of iron, magnesium and silicon, it is dense, hot and semi-solid. Earth’s crust is extremely thin. The crust, highly variable in its thickness, up to 100 km, is made of relatively light elements, especially silica, aluminum and oxygen. It’s also highly variable in its thickness. Along with the upper zone of the mantle, the crust is broken into big pieces, tectonic plates. The radius of the Earth is 6371 km, with an average density of $5.5\,g/cm^3$. The density of the inner ’solid’ core is between $9.9\div 12.2\,g/cm^3$ and the outer core’s density is between $12.6 \div 13\,g/cm^3$. The density profile in the Earth in a sphere of radius R can be visualized in the paper of Sofue \cite{Sofue:2020mda}.

\section{Macroscopic composites as DM} 
\label{sec:macro}

Macroscopic dark matter (usually called macros) represents a wide class of alternatives to particle dark matter –- in fact large objects, supposed to be probably composites of fundamental particles.
Despite the fact that for macros the dominant interactions can be strong, are ``dark systems'' because their large masses imply a low number density and a small geometric cross-section per unit mass, even though the cross-section of each object is large. There remains a large range of macro mass $M_X$ and geometric cross-section $\sigma_X$ that is still not probed by experiments or observations. In these cases, the macros are quarks or baryons bound in accordance with usually fundamental forces. Recently, attention has been paid to macroscopic dark matter, as primordial black holes – see for example: \cite{Green:2024bam}, \cite{Baldes:2023rqv},  \cite{Davies:2024ysj}, or $\mu$-black holes \cite{Lazanu:2020qod} and to macroscopic composite objects, as nuclearites \cite{DeRujula:1984axn} and strangelets: \cite{Alcock:1985vc}, \cite {Farhi:1984qu}, \cite{Lynn:1989xb}. 
These macroscopic objects made of baryons may be stable with sufficient strangeness. A distinct idea is associated with the existence of nuggets, originally started with the work of Witten \cite{Witten:1984rs} and that were extended in 2003 by Ariel Zhitnitsky considering the existence of axions in the structure of these nuggets \cite{Zhitnitsky:2002qa}.  The axion quark nuggets model represents a viable model for dark matter, that is in accord with current hypothesis of the cosmology. In a series of very recent papers, Zhitnitsky and co-workers suggest that this model is able to explain very different phenomena \cite{Majidi:2024mty},  \cite{Sommer:2024},  \cite{Sekatchev:2025ixu}, \cite{Majidi:2025ylh}, for example.

\subsection {$AQN$ and $\rm{A\bar{Q}N}$}
Axion Quark Nuggets ($AQN$) and axion anti-Quark Nuggets ($\rm{A\bar{Q}N}$) are composite objects with (anti)baryons in the colour superconducting phase squeezed by the axion domain wall as the shell and considered as a new form of macroscopic candidate for dark matter. Regarding the structure, inside the AQN /A(anti)QN there exist several regions with distinct length scales. 

\textbf{Internal core.}
The central core represents the size of the nugget filled by antiquarks in a superconducting phase with interactions dominated by QCD theory. Following the predictions of this theory, it is expected that the characteristic length scale $R_{QCD}$ is approximately $10^{-13}$ cm. This length is further extended in accord with the baryonic number as $R = R_{QCD}B^{1/3}$.

Two phases are usually considered and correspond to different values of the electric charge for identical baryonic number $B$. 
In the colour--flavour locking (CFL) phase the charge scales as
\begin{equation}
Z \simeq -0.3\,B^{2/3} \simeq -(30 \div 3\times 10^{16}),
\end{equation}
for $B$ in the interval $10^{3} \leq B \leq 10^{25}$.
This range of values is very different in the case of the 2CS phase, for which
\begin{equation}
Z \simeq -5\times 10^{-3}\,B \simeq -(5 \div 5\times 10^{22}),
\end{equation}
for the same interval of baryon numbers.
In the absence of experimental constraints, most works in the literature adopt the 2CS phase.

\textbf{Colour-flavour phases.}
In the colour--flavour locking phase at high density QCD~\cite{SonStephanov2000}, quark excitations exist and are separated by the superconducting (BCS) gap $\Delta$, while gluons acquire a mass of order $g_\mu$ (where $\mu$ is the quark chemical potential) due to the Meissner effect.
The meson masses are significantly smaller than in the normal QCD vacuum and depend only on the bare quark masses.
Using $m_u = 4~\mathrm{MeV}$, $m_d = 7~\mathrm{MeV}$, and $m_s = 150~\mathrm{MeV}$, the resulting spectrum of light mesons is
\begin{align}
m_{\eta} &= 117~\mathrm{MeV}, &
m_{\eta'} &= 30~\mathrm{MeV}, \nonumber\\
m_{\pi^{0}} &= 53~\mathrm{MeV}, &
m_{\pi^{\pm}} &= 53~\mathrm{MeV}, \nonumber\\
m_{K^{0}} &= 76~\mathrm{MeV}, &
m_{K^{\pm}} &= 72~\mathrm{MeV}.
\end{align}
The mixing of the neutral pion with the $\eta$ meson occurs at the level of a few percent, while the $\eta$--$\eta'$ mixing can reach approximately $20\%$.
These differences in hadronic masses are expected to affect the annihilation processes occurring in the nugget core.

The baryonic mass component of an antiquark nugget is estimated as
\begin{equation}
M_{\rm AQN} \simeq m_p B \simeq 16.7 \left(\frac{B}{10^{25}}\right)\,\mathrm{g}.
\end{equation}
The corresponding radius is
\begin{equation}
R_{\rm AQN} = \left(\frac{3 M_{\rm AQN}}{4\pi \rho_n}\right)^{1/3}
\simeq 2.25 \times 10^{-5}\,\mathrm{cm},
\end{equation}
where the nugget is assumed to be in the CS phase and the nuclear density is taken as
$\rho_n = 3.5 \times 10^{14}\,\mathrm{g\,cm^{-3}}$.

In the atmosphere, characterized by number density $n$, the typical internal temperature of the nugget is~\cite{Zhitnitsky2025}
\begin{equation}
T \sim 20~\mathrm{keV}
\left(\frac{n}{10^{21}\,\mathrm{cm^{-3}}}\right)^{4/17}
\left(\frac{\kappa}{0.1}\right)^{4/17},
\end{equation}
where $\kappa$ is a phenomenological coefficient in the range $0 < \kappa < 1$.
In high--density media, such as rock, the temperature of an anti--quark nugget is found to lie in the interval
$T \simeq 80$--$90~\mathrm{keV}$~\cite{Flambaum2025}.

\textbf{Electrosphere.}
To maintain the overall charge neutrality of the $\rm{A\bar{Q}N}$, a cloud of positrons exists around the core. The electrosphere also plays a significant role in the $\rm{A\bar{Q}N}$ interactions with the surrounding environment. In particular, the positron cloud is responsible for the thermal radiation emitted by AQNs when they interact with ambient matter.

For weakly bound positrons from the electrosphere, the number of positrons that leave the system can be estimated as
\begin{equation}
Q \simeq \frac{4\pi R^{2}}{\sqrt{2\pi\alpha}}\,(m T)\left(\frac{T}{m}\right)^{1/4}
\simeq 1.5\times 10^{6}
\left(\frac{T}{\mathrm{eV}}\right)^{5/4}
\left(\frac{R}{2.25\times 10^{-5}\,\mathrm{cm}}\right)^{2},
\end{equation}
where $R$ is the nugget radius, $T$ is the temperature of the positron cloud, $m$ is the electron mass, and $\alpha$ is the fine-structure constant. The typical thickness of the electrosphere is of the order of $10^{-8}\,\mathrm{cm}$.

\textbf {Axion domain wall.}
The total energy of an  $AQN$ or  $\rm{A\bar{Q}N}$ reaches its equilibrium minimum when the axion domain wall contributes approximately one third of the total mass. In this equilibrium configuration, the system does not emit axions. However, this situation changes when part of the baryon charge of the $\rm{A\bar{Q}N}$ is annihilated due to interactions with the environment. In this case, the $\rm{A\bar{Q}N}$ becomes unstable with respect to axion emission, since it is no longer in its minimum-energy configuration.

As a consequence of baryon annihilation, the $\rm{A\bar{Q}N}$ loses mass and its size shrinks, causing the axion domain wall to oscillate. These oscillations excite internal modes of the domain wall and ultimately lead to the radiation of propagating axions. The thickness of the domain wall is determined by the ALP mass and is of the order of the ALP reduced Compton wavelength $\lambda$, given by~\cite{Afach:2021}
\begin{equation}
\Delta x \simeq 2\sqrt{2}\
 \frac{\hbar}{m_a c},
\end{equation}
where $m_a$ is the ALP mass. In ref.~\cite{Afach:2021}, the numerical prefactor is obtained by approximating the spatial profile of the field-gradient magnitude as a Lorentzian and defining the thickness as the full width at half maximum (FWHM).

The duration of the ALP emission signal is then
\begin{equation}
\delta t = \frac{\Delta x}{v_\perp}
= \frac{2\sqrt{2}\,\hbar}{v_\perp m_a c}
= \frac{2\sqrt{2}\,\hbar c}{v_\perp m_a c^2}
\simeq \frac{555.54\times 10^{-15}\,\mathrm{MeV\,m}}{v_\perp E_a},
\end{equation}
 For ALPs produced with velocities of order $v_\perp \simeq 0.6\,c$ and masses $m_a \gtrsim 2.15\,\mathrm{keV}$, the emission duration is of the order of $10^{-19}\,\mathrm{s}$.

\textbf {AQN-induced axions.}
The mechanism of axion production by AQNs interacting with the Earth was developed by Zhitnitsky and collaborators, see for example refs.~\cite{Ge:2018,Zhitnitsky:2021}. In this framework, a key feature of the emitted axions is that they are produced with relativistic velocities, typically $v_{A\bar{Q}N} \simeq 0.6\,c$.

Depending on the axion mass, which is currently unknown, the axion emission region can extend to distances of order centimeters~\cite{Sommer:2024}. Experimental searches for axions have covered more than 20 orders of magnitude in mass. In the present work, we focus on axion-like particles (ALPs) and assume that their dominant coupling is to electrons.

In the cited literature and references therein, several possible interaction scenarios involving the different components of the AQN, or the nugget as a whole, are discussed. We do not repeat these analyses here.

\subsection{Possible explanations of the unobservability of dark matter}
\label{sec:DM_unobservability}

The rates of interactions for a generic type of dark matter, $X$, are proportional to the product between $n_X \sigma_X v$, (the matter number density, the interaction cross section and velocity (or relative velocity). Using $n_X = \rho_X / M_X$, it appears clear that the rate of interaction will be proportional with the ratio $\sigma_X / M_X$ with $(\rho_X v)$ where the ratio $\sigma_X / M_X$ is essential if the product $\rho_X v$ is approximately constant. Thus, two possibilities appear: a) the non-observability of the dark matter is the consequence of the small value of the cross section, when the dominant dark matter interaction is weak as in the case of WIMPs or other predicted components of the DM, or, b) interactions of DM can be dominated by strong interaction, but their masses are very heavy and the ratio $\sigma_X / M_X$ is a very small number.
The latter possibility was discussed early on in ref.~\cite{Starkman1990} and more recently in the context of macroscopic dark matter in ref.~\cite{Jacobs2015}.

WIMPs constitute a well-studied class of candidates with masses typically in the range
$10~\mathrm{GeV} \lesssim M_{\rm WIMP} \lesssim 10^{4}~\mathrm{GeV}$.  
Recent results from the XENONnT experiment~\cite{XENONnT2023} exclude spin-independent cross sections above
$2,58 \times10^{-47}~\mathrm{cm}^2$ at $M_{\rm WIMP}=28~\mathrm{GeV}$ (90\% C.L.), while the LUX-ZEPLIN experiment~\cite{LZ2022}
excludes cross sections above $9,2 \times 10^{-48}~\mathrm{cm}^2$ at $M_{\rm WIMP}=36~\mathrm{GeV}$ (90\% C.L.).
These limits correspond to
\begin{equation}
\frac{\sigma_{\rm WIMP}}{M_{\rm WIMP}} \sim 10^{-25}~\mathrm{cm}^2\,\mathrm{g}^{-1}.
\end{equation}

Strongly interacting massive particle scenarios were also analyzed in ref.~\cite{Starkman1990}.  
More recently, axion quark nuggets (AQNs) were proposed from a similar perspective in ref.~\cite{Sidhu2020}.
For macroscopic DM candidates, the geometric cross section can be estimated as
\begin{equation}
\sigma_X \simeq 2 \times 10^{-10}
\left(\frac{M_X}{\mathrm{g}}\right)^{2/3}
\mathrm{cm}^2 .
\end{equation}
For a typical AQN mass $M_{\rm AQN} \simeq 10^{24}~\mathrm{GeV} \simeq 1.67~\mathrm{g}$, one finds
\begin{equation}
\frac{\sigma_{\rm AQN}}{M_{\rm AQN}} \sim 10^{-10}~\mathrm{cm}^2\,\mathrm{g}^{-1},
\end{equation}
which is many orders of magnitude larger than in the WIMP case, yet still compatible with non-observation due to the large mass scale.

\subsection{Total mass of dark matter captured by the Earth: simple estimates}
\label{sec:DM_capture_Earth}

We assume that the Solar System is embedded in a dark matter halo characterized by a Maxwellian velocity distribution and an approximately uniform density.
According to ref.~\cite{Bertone2005}, the local Galactic DM density is
\begin{equation}
\rho_G \simeq 4 \times 10^{-25}~\mathrm{g\,cm^{-3}},
\end{equation}
with an upper bound on the DM density within the Solar System of
\begin{equation}
\rho_{\rm SS} < 4.4 \times 10^{-19}~\mathrm{g\,cm^{-3}},
\end{equation}
as inferred from planetary dynamics~\cite{Khriplovich2007}.

The capture of DM by the Earth requires, at minimum, a three-body gravitational interaction involving the Earth, the Sun, and the DM particle, thus DM is captured in the gravitational field of the two heavy ones.
Such processes were studied in refs.~\cite{Khriplovich2009,Khriplovich2011}.  
These works estimate that the DM density captured by the Earth is of order
\begin{equation}
\rho_{\rm capt} \simeq 9.3 \times 10^{-22}~\mathrm{g\,cm^{-3}},
\end{equation}
leading to a total accumulated DM mass of approximately
\begin{equation}
M_{\rm capt}^{\oplus} \sim 4 \times 10^{18}~\mathrm{g}.
\end{equation}

\subsection{The effect of the gravitational field of the Earth}
\label{sec:grav_earth}

The gravitational field of the Earth modifies the trajectories of $\rm{A\bar{Q}Ns}$ in its vicinity, as originally suggested by Dyson~\cite{Dyson1963}. More recently, gravitational focusing of low-velocity dark matter near the Earth has been discussed by Sofue~\cite{Sofue:2020mda} and by Kryemadhi \textit{et al.}~\cite{Kryemadhi2023}. 

Flambaum and collaborators~\cite{Flambaum2025} argued that the Earth's gravitational field is too weak to significantly accelerate an anti-quark nugget approaching from infinity, and that the size of the Earth is insufficient to fully stop $\rm{A\bar{Q}N}$s. According to their analysis, only $\rm{A\bar{Q}N}$s with baryon number \begin{equation} A \lesssim 10^{23} \end{equation}
can be stopped inside the Earth or gravitationally captured. $\rm{A\bar{Q}N}$s with \begin{equation} A \gtrsim 10^{24} \end{equation} are expected to traverse the Earth, losing less than $10\%$ of their mass and escaping gravitational capture.

This assumption is not adopted in the present work. Instead, we assume a plausible accumulation of $\rm{A\bar{Q}Ns}$ in the Earth's core, motivated by the expected enhancement of their density in the central region due to repeated interactions and long-term gravitational effects.

\section{Axions produced by $\rm{A\bar{Q}N}$} 

\subsection{Axion and ALP properties: masses and effective coupling constants}
\label{sec:alp_properties}

Some discussion of the expected properties of axions is useful. The Standard Model contains a source of CP violation from QCD dynamics, which has been observed. A most attractive solution is the Peccei--Quinn (PQ) mechanism~\cite{PecceiQuinn1977}, which predicts the existence of a QCD axion - see papers of Weinberg and Wilczek ~\cite{Weinberg1978,Wilczek1978}. These axions acquire mass through their coupling to the QCD condensate.

The interaction of axions with matter is characterized by effective coupling constants to photons, electrons, and nucleons, as well as by the PQ symmetry-breaking scale $f_a$. The axion mass $m_a$ and $f_a$ are related through
\begin{equation}
m_\pi f_\pi \simeq m_a f_a,
\end{equation}
leading to
\begin{equation}
m_a \simeq 5.69~\mathrm{eV} \left( \frac{10^6~\mathrm{GeV}}{f_a} \right).
\end{equation}

Originally, PQ symmetry breaking was assumed to occur at the electroweak scale, $f_a \simeq (\sqrt{2}G_F)^{-1/2} \simeq 250~\mathrm{GeV}$. Axion-like particles (ALPs) are hypothetical light pseudo-Nambu Goldstone bosons which do not necessarily address the strong CP problem and appear in the spontaneous breaking of a global symmetry. ALPs can interact with all particles of the SM. Their masses and coupling strengths are theoretically free parameters and can span many orders of magnitude. In certain regions of parameter space, ALPs can be non-thermal candidates for Dark Matter or, in other regions where they decay, mediators to a dark sector. For large symmetry breaking scales, the ALP can be a harbinger of a new physics sector at such high energy scales that would otherwise be experimentally inaccessible. 
Despite the possibility for ALPs to interact with a wide range of particles, most models propose interactions with photons and electrons, considering the effective Lagrangian of the form:

\begin{equation}
\mathcal{L}_{\mathrm{int}} =
-\frac{1}{4} g_{a\gamma} a F^{\mu\nu} \tilde{F}_{\mu\nu}
+ g_{ae} a \bar{e}\gamma^5 e ,
\end{equation}
where $a$ denotes the ALP field, $F^{\mu\nu}$ is the electromagnetic field strength tensor, $\tilde{F}_{\mu\nu}$ its dual, and $g_{a\gamma}$ and $g_{ae}$ are the axion--photon and axion--electron couplings.

Usually, ``true axion'' models address the problem of CP violation. 
The difference between ALPs (Axion-Like Particles) and the true axion 
lies in the fact that ALPs are not intended to solve the strong CP problem; 
otherwise, their phenomenology is quite similar.

If the confining group is larger than QCD, the relation
\begin{equation}
m_a^2 f_a^2 = \text{const.}
\end{equation}
holds, with a larger constant value. In this case, the parameter space of the 
true axion, as well as the constraints on the axion mass, become relaxed. 
Starting from ALP frameworks, the CP problem can then be addressed within 
different specific models. 

See, for example, the presentation by B.~Gavela at the Planck 2019 conference,
``On axions and ALPs''~\cite{Gavela2019}, and the references cited therein. 
Within this approach, a much wider window for the exploration of axion masses 
opens up.

In this work, we focus primarily on the axion--electron coupling, motivated by possible detection in noble-gas detectors.

Recent constraints on $g_{ae}$ were reported by Gavrilyuk \textit{et al.}~\cite{Gavrilyuk:2022yli}, using resonant excitation of the $9.4~\mathrm{keV}$ level of ${}^{83}$Kr by solar axions. Their results exclude masses below $4.6~\text{eV}$ in the DFSZ model (with $\cos^2\beta = 1$) and upper $320~\text{eV}$ in the KSVZ (for $E/N = 8/3$) respectively. In the DFSZ model, the dimensionless coupling constant $g_{ae}$ is expressed in terms of the parameter $f_a$ and free parameter $\cos^2\beta$, as
$g_{ae} = \frac{1}{3}\,\frac{m_e}{f_a}\,\cos^2\beta$,
with $m_e$ the electron mass. For the maximum value for angle,
$g_{ae} = 2.99 \times 10^{-11}\, m_a~[\text{eV}]$.
In the KSVZ model, the axion does not interact directly with electrons and the effective axion--electron constant is calculated for one-loop correction.

Saha and co-workers~\cite{Saha:2025} using the latest NIRSpec IFU spectroscopic observations from JWST, put the strongest bound on the photon coupling for QCD axion as ALP in the mass range between $0.47$ and $2.55~\mathrm{eV}$ and Kar, Roy and Sarkar~\cite{Kar:2025} analyzing multi-wavelength data obtained from the central region of Messier~87 galaxy by several telescopes: \textit{Swift}, \textit{Astrosat}, \textit{Kanata}, \textit{Spitzer} and \textit{International Ultraviolet Explorer} in the infrared to ultraviolet frequencies ($\sim 2 \times 10^{14}~\mathrm{Hz} - 3 \times 10^{15}~\mathrm{Hz}$), constrain the axion ALP mass in the range $2~\mathrm{eV} \lesssim m_a \lesssim 20~\mathrm{eV}$ in the M87 halo and in the range $8~\mathrm{eV} \lesssim m_a \lesssim 20~\mathrm{eV}$ if the two-photon coupling is considered. Yu-Xuan Chen et al.~\cite{Chen:2024} have improved the limits by up to $2$ orders of magnitude on the ALP-photon coupling in ALP mass range between $0.25~\mathrm{keV}$ to $5~\mathrm{keV}$, if the ALP-electron coupling is $g_{aee} \le 10^{-15}$. Recently, NEON Collaboration reported new constraints on ALPs using data from this experiment~\cite{PhysRevLett.134.201002}. These limits probe previously unexplored regions of the ALP parameter space. For ALP-photon coupling ($g_{a\gamma}$), limits reach as low as $6.24 \times 10^{-6}~\mathrm{GeV}^{-1}$ at $m_a = 3.0~\mathrm{MeV}/c^2$, while for ALP-electron coupling ($g_{ae}$), limits reach $4.95 \times 10^{-8}$ at $m_a = 1.02~\mathrm{MeV}/c^2$.

Several laboratory experiments: LUX, PandaX, XENON1T, Majorana, SuperCDMS, XMASS, EDELWEISS, GERDA, have published limits on ALPs with masses in the keV--MeV range couplings to electrons. Two recent papers,
\cite{PhysRevLett.128.221302,PhysRevD.109.075044}, analysed these results and imposed new constraints. Here we retain the following conclusions: these experiments can probe the ALP electron coupling in the keV--MeV mass range if
$|g_{ae}| \geq 10^{-13}$.
This very short review shows that ALP electron coupling and the masses for ALP are still in an ambiguous domain.

ALP interactions with electrons allow several detection channels. In this work, we focus on:
\begin{enumerate}
\item the axio-electric effect:
$a + (Z,e) \rightarrow e + (Z,e)^{*}$,
\item inverse Compton scattering:
$a + e \rightarrow \gamma + e$.
\end{enumerate}

Axio-electric effect is considered in analogy with the photo-electric effect, the axion is absorbed by the atom and at least, an electron is moved in an excited state. The cross section is proportional with the photoelectric cross section of the target and thus is proportional to $Z^5$, with $g_{ae}^2$ and inverse proportional with normalized velocity of the ALP. The total cross section for inverse Compton process, is dominantly proportional with $g_{ae}^2$ and for $m_a \lesssim 2~\mathrm{MeV}$, the phase-space contribution to the cross section is approximately independent of the axion mass and the cross section becomes
\begin{equation}
\sigma_{\rm IC} \simeq 4.3 \times 10^{-25} g_{ae}^2~[\mathrm{cm}^2],
\end{equation}
and multiplied with $Z$ for inverse Compton process at atomic level. The cross section for this channel is ineffective for axions with $E_a \gtrsim 50~\mathrm{keV}$~\cite{PhysRevD.111.043021}.

Using the same reference, the cross section for the axio-electric effect can be calculated supposing that the axion is absorbed by the atom and an electron with an energy equal to the difference between the axion energy and the electron binding energy will be emitted. The cross section is thus:
\begin{equation}
\sigma_{ae} = \sigma_{pe}\,
\frac{g_{ae}^2}{\beta_a}
\frac{3 E_a^2}{16 \pi \alpha m_e^2}
\left(1 - \frac{\beta_a^{2/3}}{3}\right).
\end{equation}
Here, due to presence of the photoelectric cross section $\sigma_{pe}$, the axio-electric effect is proportional to $Z^5$, and represent the main contribution for detection in elements with high-$Z$.

Using the constrain for $g_{ae} \geq 10^{-13}$, at atomic level,
\begin{equation}
\sigma_{\rm IC}~[\mathrm{cm}^2] \approx 4.3 \times 10^{-25} Z g_{ae}^2,
\end{equation}
and
\begin{equation}
\sigma_{ae}~[\mathrm{cm}^2] \approx \sigma_{pe}(Z) \times 0.4 \times 10^{-26} E_a^2~[\mathrm{MeV}].
\end{equation}

In Ar, $\sigma_{\rm IC} \sim 7.7 \times 10^{-50}~\mathrm{cm}^2$ and $\sigma_{ae} \sim 4 \times 10^{2}~\mathrm{cm}^2$ (if the energy of the axion is considered around $2.9~\mathrm{keV}$ and cross section for photoelectric effect is considered from~\cite{XCOM}).

For the number of ALPs from AQ$\bar{\mathrm{N}}$s, we will use the assumptions introduced by Ariel Zhitnitsky and collab.~\cite{Zhitnitsky:2021,Sommer:2024}, as well as the references cited in these papers.
Starting with the hypothesis of zero baryon net charge in the Universe, they predict that the observed dark-to-visible-matter-density ratio is
\begin{equation}
\Omega_{\rm dark} : \Omega_{\rm visible} \simeq 5:1 ,
\end{equation}
and this implies a baryon charge ratio of
\begin{equation}
B_{\bar{A}{\rm QN}} : B_{{\rm AQN}} : B_{\rm visible} \simeq 3:2:1 ,
\end{equation}
where the subscripts of $B$ correspond to the baryon charges of antimatter AQ$\bar{\mathrm{N}}$s, matter AQNs, and visible matter respectively.

A typical value for the AQN mass is
\begin{equation}
M_{\rm AQN} \approx 16~{\rm g}\times \left(\frac{B}{10^{25}}\right)
\end{equation}
see \cite{Zhitnitsky2024}.
For a baryonic number $B=10^{24}$, the maximum number of AQNs accumulated in the core of the Earth is approximately
\begin{equation}
\sim \frac{4\times 10^{18}~{\rm g}}{1.6~{\rm g}} \sim 2.5\times 10^{18}.
\end{equation}
Using the predictions of the axion generation model in AQN, the mass of the domain wall possible to be converted into axions is
\begin{equation}
M_{\rm AQN}^{\rm axions} \approx 5.3~{\rm g}\times \left(\frac{B}{10^{25}}\right),
\end{equation}
while the mass of the AQN core able to annihilate is
\begin{equation}
M_{\rm AQN}' \approx 10.7~{\rm g}\times \left(\frac{B}{10^{25}}\right).
\end{equation}

Depending on the axion mass range considered, the number of axions produced (for $B=10^{24}$) is as follows.
For axion masses in the range
\begin{equation}
10^{-6}~{\rm eV} < m_a < 10^{-2}~{\rm eV},
\end{equation}
the number is between $0.3\times 10^{39}$ and $0.3\times 10^{35}$~\cite{Ge:2018}.
If the axions are considered as ALPs, with masses in the range
\begin{equation}
0.47~{\rm eV} \lesssim m_a \lesssim {\rm MeV},
\end{equation}
the maximum number of axions produced is between 
\begin{equation}
0.6\times 10^{33} \; \rm{and}\; 3\times 10^{26}\; \rm{per} \; A\bar{Q}N.
\end{equation}

The next aspect concerns whether these particles can survive up to an adequate detector, or convert or decay into other detectable particles.
Usually, it is expected that axions penetrate the structure of the Earth without a significant number of interactions.
For example, in the Sun, the mean free path is
\begin{equation}
\lambda \approx 5\times 10^{18}~{\rm cm}
\end{equation}
for 4~keV axions~\cite{Battesty2008}.
For $\rm{A\bar{Q}N}$-induced axions, the electromagnetic signal is very broad, in contrast with the narrow line (with $\Delta\nu/\nu \sim 10^{-6}$) searched for in galactic axion experiments~\cite{Liang2021}.

The tree-level Feynman diagrams for the inverse Primakoff process on electrons or nucleons, the inverse Compton process, or bremsstrahlung radiation are not adequate, because the final photon, electron, or nucleon cannot survive up to underground detectors.
If the associated Lagrangian for such particles contains terms of the form $\phi^4$, with $\phi$ being the field representing the particles, this opens the possibility of a connection with neutrino physics.
Civitarese and collaborators~\cite{Civitarese2025,Civitarese2023} claimed the possibility of the existence of a mechanism for axion--neutrino couplings.
A detailed analysis of the elastic process $a+\nu$ was performed by Reynoso \textit{et al.}~\cite{Reynoso2022}.

Axions are expected to penetrate the Earth with negligible attenuation.
For $\rm{A\bar{Q}N}$ induced axions, the electromagnetic signal is very broad, in contrast to the narrow spectral line
($\Delta \nu / \nu \sim 10^{-6}$) expected for galactic axions \cite{Liang2021}.

Tree-level processes such as the inverse Primakoff effect, inverse Compton scattering, or bremsstrahlung are inefficient
because the final-state photons or charged particles cannot propagate to underground detectors.
If the effective Lagrangian contains $\phi^4$ terms, with $\phi$ denoting the axion field, couplings to neutrinos become
possible. Such axion--neutrino interactions were discussed in
\cite{Civitarese2025,Civitarese2023}, and elastic $a+\nu$ scattering was analyzed in detail in \cite{Reynoso2022}.

\section {Detection of the axions produced by $\rm{A\bar{Q}Ns}$}
\subsection{Can axions be detected using liquid noble gases?}

Liquefied noble gases are well-suited detector media for these rare event search because present a very efficient scintillation (fast and with high light yields). Ionizing particles produces in liquid argon/xenon excitations and ionizations. In the excitation processes, dimers are excited and occur in singlet and triplet states with very different lifetimes, about 6 ns and about 1.5 microseconds, respectively. The linear energy transfer (LET) determines the relative population of these states; the triplet states with long lifetimes are produced at high LET, the singlet state has a much shorter lifetime, and thus it is possible to discriminate very effectively between heavy particles interactions and interactions with low LET, such as low energy electrons or gammas, by digitizing pulses and comparing the light emission in the first tens of nanoseconds and the full pulse duration.

The mechanism of light emission after excitation or single ionization by an incident particle can be discussed as a process with three possibilities.
\begin{itemize}
\item[a)] After the collision between incident particle and A atom, if a single ionization is produced, and in this process an ionic excimer ($\mathrm{Ar}_2^+$) is formed; usually, excimers recombine with an electron and double excited atom is produced ($\mathrm{Ar}^{(**)}$).
\item[b)] The incident particle produced multiple ionizations ($\mathrm{Ar}^{(x*)}$) that in similar evolutions as in previous case produces resonance lines and / or third continuum.
\item[c)] Last mechanism is the excitation ($\mathrm{Ar}^{(**)}$). In interaction with another Ar will be produced de-excitation in a lower state ($\mathrm{Ar}^*$) followed by the excimer formation in interaction with an atom.
\end{itemize}

The excimers are the sources for left turning point, first and second continuum in de-excitations spectrum. Martin Hofmann~\cite{Hofmann2012} suggest the following radiatively decay of excimer molecules: 188~nm ($\mathrm{Ar}_2^{2+}$), 128~nm ($\mathrm{Ar}_3^{2+}$), 177~nm, 212~nm, 225~nm ($\mathrm{Ar}_2^{+*}$), 245~nm ($\mathrm{Ar}_3^{+*}$).

Using the relativistic Hartree-Fock formalism,~\cite{Dzuba2010} calculated the rates of atomic ionization in Ar and Xe for mass range relevant for dark matter searches. In the investigations of the energies necessary to produce excitation and corresponding emission processes, important is the argon excimer molecule potentials, that are dependent on the intermolecular distances. A very intuitive figure was presented by Neumeier~\cite{Neumeier2015Thesis}; see figure~1.6 from the cited paper.

The molecule potentials are dependent by the internuclear distance. The ground state potential is repulsive. Two upper curves correspond to the potential of a singlet excimer formed by a neutral Ar atom in the state (${}^1S$) and one excited (${}^3P_2$), the singlet excimer: $\mathrm{Ar}_2^*({}^1\Sigma_u^+)$ and above the triplet excimer $\mathrm{Ar}_2^*({}^3\Sigma_u^+)$ formed also by a neutral atom and one excited in a triplet state: $\mathrm{Ar}({}^1S)+\mathrm{Ar}({}^3P_1)$. At low collision rates, corresponding to large internuclear distances, resonance emission can be observed. At short distances formation of excimer molecules in different vibrationally excited states are possible and thus the decay of the excited excimers states produced first and second continuum, as well as the classical left turning point.

The next point in this analysis is the comparison between gaseous and liquid states. Besides the second excimer continuous which is the domination the spectrum and is similar in both phases of argon and the first excimer continuum at shorter wavelengths -- difficult to be separated, exist several distinct behavior in the emission features: in liquid argon LTP is a very tiny intensity (observable at 155~nm); beginning from approximatively 170~nm the third excimer continuum arise (produced by ionization and not by excitation), is suppressed compared to the second and shows two maxima instead of one. In pure liquid argon was put in evidence a strong emission features in the near-infrared (around 950~nm).

Another option for detector is Xe doped LAr. The mechanisms for excimers formation in xenon are similar with the case of argon, but with different distances in the energy levels. In xenon, the dominant transition is also second continuum, at $\simeq174$~nm. Neumeier et~al.~\cite{Neumeier2015EPL} investigated light emission from xenon-doped liquid argon. The transfer almost saturates at a xenon concentration of 10~ppm for which, in addition, an intense emission in the infrared at a peak wavelength of 1173~nm, with $13000 \pm 400$ photons per MeV deposited by electrons had been found, following reaction:
\begin{equation}
\mathrm{Ar}_2^*\left(({}^{1,3})\Sigma_u\right) + \mathrm{Xe}({}^1S_0)
\rightarrow
2\,\mathrm{Ar}({}^1S_0) + \mathrm{Xe}({}^1P_1).
\end{equation}

Considering only kinematic aspects, the maximum energy transfer to an electron in a single binary collision of an axion is around 10~eV, for values of the mass, $m_a \ge 2.15$~keV, all the possibilities to excitations in LAr, LXe or Xe doped LAr are satisfied and thus scintillations are possible. This mass constraint is in accord with present limits imposed by all recent searches. Thus, this very simple mechanism can be used for detection.

Heaton et. al. \cite{heaton2026probingkevmassqcd} probing keV mass QCD axions in the range 3,46 - 3,48 keV with the SACLA X-ray free electron laser, claim that their results are the most stringent laboratory constraints in this mass range for both KSVZ and DFSZ models, in accord with the results predicted by the present work.

\subsection{Expected event rates and background considerations}

The number of events produced by axions from a single $\rm{A\bar{Q}N}$ interacting with target nuclei is
\begin{equation}
N_{\mathrm{events}} \simeq
\Phi \times \sigma \times N_{\mathrm{target}} \times \Delta t ,
\end{equation}
where $\Phi$ is the number of incident particles passing through a unit area and unit time,
$N_{\mathrm{target}}$ is the total number of electrons in the target material that can participate in the process,
and $\sigma$ is the cross section for the specified interaction.
The flux is given by
\begin{equation}
\Phi = \frac{N_{\mathrm{ALP}}^{\mathrm{detector}}}{S \times \delta t},
\end{equation}
where $S$ is the surface of the detector, $\delta t$ is the time interval during which axions are emitted, and
$\Delta t$ is the time to produce the signal in the detector (time to generate the scintillation signal).

For some elementary estimations, we consider the parameters of one module from the DUNE far detector filled with Xe-doped LAr,
supposing the predicted constraint for the axion mass to be $\gtrsim 2.15~\mathrm{keV}$, their interactions in argon being
dominated by the axio-electric effect with a typical cross section of the order of
$10^{-48}~\mathrm{cm}^2$ and a fast time of de-excitation in the scintillation between picoseconds and nanoseconds.
For an A$\bar{\mathrm Q}$N with $B=10^{25}$, the number of produced axions is $1.4\times10^{30}$ and the number of scintillations
produced (events) is between $1.4\times(10^3\div10^6)$.

In the detection, a first trigger in the searches of signals associated with the ALP interactions consists in the selection
of the directionality of the particles: only particles that enter the detector through the bottom and move upwards will be
selected and with expected energies.

The technical problem is to measure these expected unusual transitions in LAr or Xe-doped LAr associated with a very small
number of optical photons. Due to the low photon number, highly sensitive detectors and a large sensitive area are required.
Two examples for potential solutions are discussed.
For example, Fischer, Keller and Ritzert~\cite{Fischer:2025} created a digital SiPM photo-detection chip combining single-photon
sensitive avalanche photodiodes and CMOS readout electronics on a single die.
The ARAPUCA (Argon R\&D Advanced Program @ UniCAmp) concept, proposed by Segreto and Machado~\cite{Machado:2016}
and further developed in~\cite{Motta:2018}, is to trap photons inside a box with highly reflective internal surfaces, so that
the detection efficiency of trapped photons is high even with a limited active coverage of its internal surface.
ARAPUCA technology inherently allows for discriminating scintillations at different wavelengths through its layered setup of
dichroic filters and wavelength-shifting materials. This allows it to isolate and detect specific components of light,
effectively discriminating between different wavelength scintillations.

A crucial problem is the capability of these detectors to discriminate between the signals of interest and the background
and the identification of other sources that mimic the expected processes.
In ton-scale to multi-kton-scale liquid detectors such as LUX-ZEPLIN (LXe), ProtoDUNE (LAr) and DUNE (LAr or Xe-doped LAr),
numerous radioactive materials are present in the surrounding environment, such as rock in the underground cavern walls,
dust, radon and daughters, as well as components within the detector.
Usually muons, neutrons and neutrinos, as primary or secondary particles, represent the main sources of cosmogenic background
for this kind of experiment.
Inside LAr detectors the intrinsic radioactive background is dominated by the content of $^{39}$Ar in natural atmospheric
argon, with an upper limit of $(8.0 \pm 0.6)\times10^{-16}~\mathrm{g/g}$, or $(1.01 \pm 0.08)~\mathrm{Bq/kg}$.
For this undesirable isotope, with a $\beta$ end-point energy of $565~\mathrm{keV}$, electrons have very short tracks
($\sim1~\mathrm{mm}$), but antineutrinos are also generated.
In this case the background value for the energy loss is $dE/dx \simeq 2.1~\mathrm{MeV/cm}$.
DUNE, for example, requires an average light yield of $>20$ photoelectrons/MeV with a minimum of
$0.5$ photoelectrons/MeV, corresponding to a photon detection efficiency (PDE) of $2.6\%$ and $1.3\%$, respectively.
For detectors placed at the surface, particularly in the case of ProtoDUNE Vertical Drift, there exists an additional
background source associated with its position with respect to the beam pipe~\cite{Parvu:2021}.

Comprehensive analyses of the backgrounds are published in
refs.~\cite{LZ:2023,Lazanu:2024}.
In Bezerra \textit{et al.}~\cite{Bezerra:2023}, the possibilities to increase sensitivity to low-energy physics in a third
or fourth DUNE-like module are also discussed, with careful controls over radiopurity and targeted modifications to a detector
similar to DUNE, thus obtaining a detector with very low background.

Detectors which utilize scintillators have a high light yield (typically producing $\mathcal{O}(10^{4})$ visible photons
per MeV of deposited energy) but are unable to reconstruct directional information from isotropic scintillation photon
emission, contrary to detectors designed for reconstruction of Cherenkov radiation, emitted at a characteristic angle with
respect to the tracks of relativistic particles and able to reconstruct directional information and discriminate between
particle types.
Having a poorer energy resolution and higher kinetic energy detection thresholds because the Cherenkov photon yield is
$\mathcal{O}(10^{2})$ smaller than that of high light-yield liquid scintillators, this technology can nevertheless be used
to evidence the signals of interest for the present paper.

Future detectors such as EOS or THEIA will utilize differences in both scintillation and Cherenkov light emission for
different particle species to perform background rejection.
Possible strategies to implement this method are discussed in
refs.~\cite{AguilarArevalo:2025,Callaghan:2023}.

\section{Summary}

Axion anti-quark nuggets provide a macroscopic dark matter candidate with strong interactions.
AQ$\bar{\text{N}}$s captured by the Earth may become unstable due to baryon annihilation, leading to axion emission from
oscillating domain walls.
For ALP masses above $\sim 2.15$~keV, interactions with liquid noble gases can induce detectable scintillation signals,
making large liquid argon and xenon detectors promising instruments for probing this scenario.


\acknowledgments
 The author (IL) is deeply thankful to Dr. Mihaela P{\^a}rvu for her continuous support.
For IL this work was supported by the Romania - CERN Program, under contract CERN-RO/CDI/2024-001/25.11.2024.


\bibliographystyle{JHEP}
\bibliography{biblio.bib}






\end{document}